\documentstyle[12pt]{article}
 
\def\0{\over } \def\1{\vec } \def\2{{1\over2}} \def\4{{1\over4}}
\def\5{\bar } 
\def\6{\partial }
\def\7#1{{#1}\llap{/}}
\def\8#1{{\textstyle{#1}}} \def\9#1{{\bf{#1}}}
\def\.{\cdot }
\def\^#1{\widehat{#1}}
 
  \let\g=\gamma \let\d=\delta
  \let\h=\eta \let\th=\theta
  \let\l=\lambda \let\m=\mu
\let\n=\nu   \let\r=\rho 
\let\t=\tau \let\o=\omega 
\let\ph=\varphi   
   
\let\L=\Lambda  \let\D=\Delta

\def\CL{{\cal L}}
\def\CM{{\cal M}}
 
\def\({\left(} \def\){\right)} \def\<{\langle } \def\>{\rangle }
\def\[{\left[} \def\]{\right]}  
 
\def\pmbf#1{\setbox0=\hbox{${#1}$}
        \kern-.025em\copy0\kern-\wd0
        \kern.05em\copy0\kern-\wd0
        \kern-.025em\raise.0433em\box0 }

\def\be{\begin{equation}}
\def\ee{\end{equation}}

\newcommand{\bel}[1]{\begin{equation}\label{#1}}
\def\bea{\begin{eqnarray}}
\newcommand{\beal}[1]{\begin{eqnarray}\label{#1}}
\def\eea{\end{eqnarray}}
\def\nn{\nonumber\\ }

\begin{document}

\begin{titlepage}
\begin{flushleft} 
TUW 97-12 \hfill hep-th/9707163 \\
ITP-SB 97-37
\end{flushleft}  
\begin{center}  \vfil 
{\large \bf 
 No saturation of the quantum Bogomolnyi bound \\
by two-dimensional supersymmetric solitons \\[12pt]
   }
\vfil 
{\large  
A. Rebhan$^1$ and P. van Nieuwenhuizen$^2$
}\\
\end{center}  \medskip \smallskip \qquad \qquad 
{\sl $^1$} \parbox[t]{12cm}{ \sl 
  Institut f\"ur Theoretische Physik, Technische Universit\"at Wien, \\
  Wiedner Hauptstr. 8--10, A-1040 Vienna, Austria\\ } \\
\bigskip \qquad \qquad 
{\sl $^2$} \parbox[t]{12cm}{ \sl 
  Institute for Theoretical Physics, S.U.N.Y. at Stony Brook, \\
  Stony Brook, NY 11794-3840, U.S.A.\\ } \\
\vfil
\centerline{\large  ABSTRACT}\vspace{.5cm}
We reanalyse the question whether the quantum Bogomolnyi bound is
saturated in the two-dimensional supersymmetric kink and sine-Gordon
models. 
Our starting point is the usual expression for the one-loop correction
to the mass of a soliton in terms of sums over zero-point energies.
To regulate these sums, most authors put the
system in a box with suitable boundary conditions, and impose an
ultraviolet cut-off. We distinguish between an energy cut-off and a
mode number cut-off, and show that they lead to different results.
We claim that only the mode cut-off yields correct results, and only if
one considers exactly the same number of bosonic and fermionic modes in
the total sum over bound-state and zero-point energies.
To substantiate this claim, we show that in the sine-Gordon model
only the mode cut-off yields
a result for the quantum soliton mass that is consistent with the exact
result for the spectrum as
obtained by Dashen {\em et al.} 
from quantising the so-called breather solution.
In the supersymmetric case, our conclusion is that contrary
to previous claims the quantum Bogomolnyi
bound is not saturated in any of the two-dimensional models considered.
\end{titlepage}

\section{Introduction and conclusion}

More than a decade ago the quantization of solitons in 1+1-dimensional models
\cite{R,DHN2,CSG,DHNSG,DV,Verw} was extended to their
supersymmetric versions \cite{DVFH}, and the problem of how to compute
the mass of a soliton at the quantum level was studied by several
authors using various methods \cite{dAdV,Sch,Rouh,KR,IM,Y,Uch,CM,U1,U2}.
With the present activity in quantum field theory regarding
dualities between extended objects and pointlike particles, interest
in the quantum mass of solitons has come back. In view of certain
discrepancies in the above-quoted literature, we reanalyse here the
question whether the Bogomolnyi bound in the supersymmetric kink and
sine-Gordon model is saturated at order $\hbar$. Since in two-dimensional
supersymmetry no ``multiplet shortening'' arises, the beautiful and
simple arguments which have been used in four-dimensional $N=2$ models
\cite{WO} do not apply, and a more detailed analysis seems necessary.

Our starting point is an expression for the order $\hbar$ corrections to
the mass of an extended object in terms of sums over zero-point energies
of fluctuations around it,
\be
M=M_{cl.}+{\hbar\02}\(\sum\o_B-\sum\o_B'\)-
{\hbar\02}\(\sum\o_F-\sum\o_F'\)+\d M \;.
\ee
Here $M_{cl.}$ is the classical mass expressed in terms of renormalized
parameters and $\d M$ denotes the counter-terms which are first determined
in the trivial vacuum by imposing a suitable set of renormalization
conditions and which we then also use in the topologically nontrivial
sector. The zero-point energies are denoted by $\hbar\omega_{B,F}$ for
bosons and fermions, respectively, and the vacuum part is marked by a prime.
The problem is then to give a precise meaning to these infinite sums.

Most authors put the system in a large box of length $L$ with suitable
boundary conditions, leading to discretized frequencies $\omega$,
and then introduce an ultraviolet cut-off $\L$ which restricts each sum
to a finite number of terms. Then one first lets $L$ tend to infinity,
and afterwards the ultraviolet cut-off is removed. There are 
several different boundary conditions on the
fluctuations that one may impose. Various authors have demonstrated
the sensitivity of the results on the boundary conditions, but (as we
discuss in the Appendix) the requirement that the zero-point energies
cancel mode by mode in the trivial vacuum together with finiteness of
the quantum mass of the soliton 
fixes this ambiguity.

We claim that the different ways of choosing
the ultraviolet cut-off that have been used in the literature
are in fact inequivalent. In particular we consider here two
schemes that are frequently adopted (although often only implicitly):
energy/momentum cut-off where all energies $\omega=\sqrt{k^2+m^2}$
in the continuum part of the spectrum are cut off at a value $\L$,
and mode-number cut-off, where in each of the four infinite sums above
the same finite number $N$ of modes is retained. Since in addition to
the continuum (scattering) states there are also in general bound
states and zero modes, we have to specify this cut-off further.
Motivated by lattice regularization \cite{DHN2,DHNSG}, we require
the total number of these modes to be the same in each sector, which in the
presence of bound states means that fewer scattering states have to
be taken into account in the nontrivial sector.

Both the energy cut-off and the mode cut-off lead to finite answers
for the quantum mass of the soliton, but they turn out to be mutually
inconsistent, as we first show in Sect.~2 in the bosonic kink model.
Some authors have tried to circumvent such issues by not considering
a finite quantization volume at all \cite{IM,Y} by using e.\ g.\
general trace formulae over the energy spectrum. It seems to us that
these approaches amout to a particular choice of the ultraviolet cut-off
(namely an energy cut-off) and thus do not resolve the ambiguity
in the procedures.
Without further input one hardly can decide which procedure
is the correct one. Turning to the sine-Gordon model in Sect.~3 we
find a similar discrepancy, but in this case we can compare the
results for the quantum soliton mass with the exact spectrum obtained
by Dashen {\em et al.} from quantising the so-called breather solution.
It turns out that the mode cut-off but not the energy-cutoff passes this
test. We then use the same mode cut-off in a calculation of the
quantum corrections to the mass of the solitons in the supersymmetric
kink and sine-Gordon systems in Sect.~4. In Sect.~5 we determine
the quantum corrections to the Bogomolnyi bound by expanding the central charge
operator to second order in the quantum fields and computing its
expectation value. 
The latter contains the same divergent expression that occurs in
the sum over zero-point energies and can be absorbed by renormalization.
Having obtained a finite result at the order $\hbar$, we can finally
answer the question whether the Bogomolnyi bound is saturated by
the quantized solitons. The answer is that it is not.

\section{Quantum corrections to the kink mass}

The simplest field-theoretical model with solitons is given by a
real scalar field in 1+1 dimensions with spontaneously broken $Z_2$ symmetry
as described by the Lagrangian
\bel{Lk}
\CL=-\2 (\6_\mu \ph)^2-{\l\04}(\ph^2-\mu_0^2/\l)^2 \;.
\ee

This model has two degenerate vacuum states, 
$\ph=
\pm\mu_0/\sqrt\l$,
and two static
stable finite-energy (soliton) solutions, the so-called ``kink''
and ``anti-kink'' \cite{R,DHN2}
\bel{Ksol}
\ph_{K,\5K}=\pm 
{\mu_0\0\sqrt\l}
\tanh\(\mu_0(x-x_0)/\sqrt2\)
\ee
with classical (unrenormalized)
rest-frame energy $M_0=2\sqrt2\mu_0^3/3\l$.

In the corresponding quantum theory we have to relate bare and renormalized
parameters through appropriate counter-terms. 
We expand $\ph$ about one of the vacua, $\ph=\mu/\sqrt\l+\eta$ where
$\mu_0^2=\mu^2+\d\mu^2$. Then
\be
\CL=-\2 (\6_\mu\eta)^2-\mu^2\eta^2-\mu\sqrt\l\eta^3-\4\l\eta^4+
\2\d\mu^2(\eta^2+2{\mu\0\sqrt\l}\eta)+O(\hbar^2),
\ee
hence the renormalized mass of the physical boson at tree-graph level
is $m^2=2\mu^2$. We fix $\delta\mu^2$ by requiring that the one-loop
tadpole vanishes exactly, which gives
\bel{dm2k}
\d\m^2=
-3i\l\hbar\int{d^2k\0(2\pi)^2}{1\0k^2+m^2}=
{3\l\hbar\02\pi}\int_0^\L {dk\0(k^2+m^2)^{1/2}}
\ee
where we have introduced an ultraviolet cutoff $\L$.

Since only the 
mass receives a divergent contribution, we may choose a minimal renormalization
scheme defined at all loops by
\bel{RSk}
Z_\l=1,\quad Z_\h=1,\quad \m_0^2=\m^2+\d\m^2
\ee

This is the renormalization scheme that has been adopted more or less
implicitly in Refs.\ \cite{R,DHN2}. It has the advantage of maximal
simplicity, but one must not forget that there are still finite
corrections if one is interested in physical definitions of the
various parameters. Defining for instance
the physical mass of the boson through the pole
of its propagator leads to an additional 
finite contribution from the self-energy
diagram with 3-vertices whereas the diagram with a 4-vertex is cancelled
by (\ref{dm2k}),
\bel{mP}
m_P^2=m^2+9\l i\hbar
\int {d^2k\0(2\pi)^2} {m^2\0(k^2+m^2)((k-p)^2+m^2)}_{|p^2\to-m^2}
=m^2-{\sqrt3\02}\hbar\l\;.
\ee

Evaluating the leading quantum corrections to the mass of the solitons
requires the computation of the functional determinant of the differential
operator describing fluctuations around the nontrivial solutions
(\ref{Ksol}).
This leads to \cite{R}
\bel{Msums}
M=m^3/3\l+{\hbar\02}\( \sum \o - \sum \o' \) + \d M(\d\m) +O(\l)
\ee
where $\o$ and $\o'$ are the eigenfrequencies of fluctuations around
a kink and the vacuum, respectively, and
\be
\d M(\d\m)=-\2\d\mu^2\int_{-\infty}^\infty\[\ph_K^2(x)-{m^2\02\l}\]dx
={m\0\l}\d\mu^2.
\ee
The latter is the contribution
to the energy of the kink induced by the counterterm $\2\d\m^2\ph^2$ in
the renormalized Lagrangian, and, as one may check, $m^3/3\l+\d M(\d\m)=M_0$.

The normal modes of fluctuations around $\ph_K$ are given by
\bel{kfl}
\(-{d^2\0dx^2}+V''(\ph_K)\)\h_n(x)=\o_n^2 \h_n(x)
\ee
and can be expressed in terms of elementary functions \cite{R}.
Eq.\ (\ref{kfl}) has two discrete eigenvalues, $\o_z=0$, corresponding to a
translational zero mode, and $\o_e=\sqrt3m/2$, which corresponds to an
excited state of the kink, followed by a continuum
of eigenvalues $\o=\sqrt{k^2+m^2}$ corresponding asymptotically to
waves with a $k$-dependent phase shift,
\bel{dk}
\h_k(x)\sim\exp\(i[kx\pm \d(k)/2]\) \quad \hbox{for $x\to\pm\infty$} \quad
\mbox{with $\d(k)=-2\arctan{3mk\0m^2-2k^2}$}\;.
\ee

In (\ref{Msums}) the difference of the two sums has to be
calculated very carefully. Both expressions are quadratically divergent in the
ultraviolet, while their
difference is still logarithmically divergent. It is advisable to
start with a finite interval $x\in(-L/2,L/2)$, $L\gg1/m$,
in order to make all eigenvalues discrete so that we are actually
dealing with proper sums.
Choosing periodic boundary conditions\footnote{Alternative boundary conditions
are discussed in the Appendix.}, $k_n$ and $k'_n$ can be labelled
by integer numbers through the relation
\bel{kdn}
k_n L +\d(k_n)=2\pi n=k_n' L \;,
\ee
and the summation in (\ref{Msums}) includes all integers $n$.

\subsection{Energy/momentum cutoff}

In the ultraviolet, 
we have already introduced a cutoff regularization when calculating the
renormalized boson mass. It seems natural to also simply cut the
diverging sums in (\ref{Msums}) such that $k_n, k_n' \le \L$. Since
both in the vacuum and in the kink sector the energy of the fluctuations
is given by the same $\sqrt{k^2+m^2}$, this corresponds to using the
same energy cutoff for the two types of fluctuations.

In the limit $L\to\infty$, the infinite sums over $n$ can be replaced
by momentum integrals using slightly different spectral densities
\be
{dn\0dk}={1\02\pi}\(L+{d\d(k)\0dk}\),\quad {dn\0dk'}={L\02\pi} \;.
\ee

This leads to
\beal{MKEMC}
M_{\mathrm EMC}&=&
m^3/3\l+\hbar\sqrt3m/4+{\hbar\02\pi}\int_0^\L dk \sqrt{k^2+m^2}\d'(k)
+{m\0\l}\d\m^2 \nn
&=& m^3/3\l+\hbar m/4\sqrt3\;,
\eea
where we have used
$$
{\hbar\02\pi}\int_0^\L dk \sqrt{k^2+m^2}\d'(k)
+{m\0\l}\d\m^2={-3m\hbar\08\pi}\int_0^{\L/m}{dx\0\sqrt{x^2+1}(x^2+\4)}
=-{\sqrt3m\hbar\06}.$$
The subscript EMC 
refers to the strict energy or momentum cutoff that we have
used here.

\subsection{Mode-number cutoff}

However, the result for the mass of the quantum kink as reported in the
literature differs from (\ref{MKEMC}). In Ref.\ \cite{DHN2}, Dashen
{\em et al.} have proposed to use a lattice regularization. On a lattice
in a finite box, the number of degrees of freedom becomes not only
countable but finite.

When comparing the contributions with and without the kink background,
we have to compare mode by mode, since in both cases the number of modes
is the same and every single mode, even when it belongs to the continuum
eventually, makes a finite contribution. Since, in the presence of the kink,
there are two discrete
modes, we have to exclude two of the
``continuum'' modes when comparing with the energy contributions of
the vacuum. Denoting by $\D E$ the contributions from the zero-point energies
we have
\beal{mnck}
\D E_{\mathrm kink}&-&\D E_{\mathrm vac}=
{\hbar\02}{\sqrt3m\02}+{\hbar\02}\sum_{-(N-1)}^{N-1}\sqrt{k_n^2+m^2}
-{\hbar\02}\sum_{-N}^{N}\sqrt{k_n'^2+m^2}\nn
&=&{\hbar\02}{\sqrt3m\02}-\hbar m+\hbar\sum_{n=1}^N
\(\sqrt{k^2_{n-1}+m^2}-\sqrt{k'^2_n+m^2}\)
\;.
\eea

Because 
\be
k_{n-1}L+[\d(k_{n-1})+2\pi]=2\pi n=k'_nL \;,
\ee
we obtain in the limit of infinite volume
\be
\D E_{\mathrm kink}-\D E_{\mathrm vac}={\hbar\02}{\sqrt3m\02}-\hbar m
-\hbar\int_0^\L{dk\02\pi}{d\sqrt{k^2+m^2}\0dk}[\d(k)+2\pi]\;.
\ee
This result differs from the corresponding result with a strict energy cutoff
by the amount
\be
-\hbar m - {\hbar\02\pi} \sqrt{k^2+m^2}[\d(k)+2\pi]\Big|_0^{\L\to\infty}=
-{3m\hbar\02\pi}\;,
\ee
where we used that $\d(k\to\infty)\to-2\pi$.\footnote{Note that
the shift in the continuum modes in (\ref{mnck}) leads to the extra
term $2\pi$ in $[\d+2\pi]$, which is essential to obtain a finite
answer. If one does not take the shift in the continuum modes into account
one gets the same result only if one (erroneously) assumes that
$\d(k\to\infty)\to0$ \cite{R}.}

The mass of the quantum kink using lattice regularization in a finite
box and consequently
a mode-number cutoff (MNC) is therefore
\bel{MKMNC}
M_{\mathrm MNC}={m^3\03\l}+\hbar m\({1\04\sqrt3}-{3\02\pi}\) \;,
\ee
or, in terms of the physical (pole) mass of the boson (\ref{mP}),\footnote{A
particular physical definition of the coupling $\l$, e.g.\
through the low-energy limit of scattering amplitudes, would also
modify the result further at order $\l^0$.}
\be
M_{\mathrm MNC}={m_P^3\03\l}+\hbar m_P\({1\0\sqrt3}-{3\02\pi}\)\;.
\ee

The difference between (\ref{MKEMC}) and (\ref{MKMNC}) can be traced
to the attractive nature of the kink. Having a fixed energy cutoff,
we are taking slightly more modes into account due to the negative
phase shift $\delta$, to wit,
\bel{MKdiff}
L{\hbar\02\pi}\int_{\L-|{\d(\L)+2\pi\0L}|}^\L dk \sqrt{k^2+m^2} 
\stackrel{L\to\infty}\longrightarrow
{\hbar\02\pi}|\d(\L)+2\pi|\L \stackrel{\L\to\infty}
\longrightarrow {3m\hbar\02\pi}\;.
\ee

In the vacuum, there is no difference between the two procedures---both
are introducing a straightforward momentum cutoff which is eventually
sent to infinity. But the mass of the quantum kink (at least it's ratio
with the physical boson mass $m_P$) is a physical quantity that should
not depend on the particular regularization procedure.

Just from the above results one hardly can decide which 
procedure is more trustworthy. But
in the following section we shall see that only the mode-number cutoff
as introduced by a finite lattice is leading to consistent results
for the spectrum of the sine-Gordon model.
Indeed, when viewed from within lattice regularization, a strict
energy cutoff appears to be a very unnatural
procedure: in the above calculation it would have meant comparing
the quantum corrections in the kink sector to those in the vacuum
by taking two slightly different lattices, the one for the kink
being a little bit finer, before taking the continuum limit. 

\section{Quantization of the sine-Gordon soliton and breather solutions}

The sine-Gordon model is defined by the Lagrangian
\bel{LSG}
\CL=-\2 (\6_\mu \ph)^2+{m_0^2\0\g} \[ \cos\(\sqrt\g\ph\)-1 \] \;.
\ee
Its potential has infinitely many degenerate minima at $\ph=2n\pi/\sqrt\g$
and gives rise to solitons (anti-solitons) which are known in closed form,
\be
\ph^{s,\5s}(x)={4\0\sqrt\g}\arctan[\exp\(\pm m_0(x-x_0)\)]
\ee
with classical (unrenormalized) rest frame energy $M_s^0=8m_0/\g$.

In the corresponding quantum field theory, expanding about the vacuum $\ph=0$,
the Lagrangian for the fluctuations becomes
\be
\CL=-\2 (\6_\mu \h)^2-\2m_0^2\h^2+m_0^2{\g\04!}\h^4-m_0^2{\g^2\06!}\h^6
+\ldots\;.
\ee
Only the seagull loops are divergent \cite{CSG,DHNSG}
so that we can choose the minimal renormalization scheme
defined by
\bel{RSSG}
Z_\h=1,\quad Z_\g=1,\quad m_0^2=m^2+\d m^2,\quad
\d m^2=\hbar m^2 {\g\04\pi}\int_0^\L {dk\0(k^2+m^2)^{1/2}}\;.
\ee

The complete counter-term is then \cite{CSG}
\be
\d \CL=-{m^2\0\g} \(e^{\d m^2/m^2}-1 \)\(1-\cos(\sqrt\l \ph)\) \;.
\ee

Because at one-loop order there is only the logarithmically divergent sea\-gull
contribution to the mass of the boson,
the above renormalized mass coincides with the physical mass defined
by the pole of the propagator, up to contributions of the order $\g^2$,
which will not be of interest to us but can be found in 
Ref.\ \cite{DV,Verw}.

\subsection{Quantum corrections to the soliton mass}

In order to compute the quantum corrections to the soliton mass, the
solutions to the analogue of eq.\ (\ref{kfl}) have to be obtained, now
with $\ph^s$ in place of $\ph_K$. This leads to a fluctuation spectrum
consisting of the translational zero mode with $\o_z=0$ and a continuum
with $\o=\sqrt{k^2+m^2}$, corresponding to two-dimensional waves with
asymptotic phase shift\footnote{Full details can be found in Ref.\ \cite{Verw}.}
\bel{dkSG}
\d(k)=2\arctan{m\0k} \;.
\ee

Proceeding as before (but setting $\hbar=1$ in what follows), we obtain
for the quantum correction to the soliton mass
\bel{MSGEMC}
\D M_s^{\mathrm EMC}=M_s^{\mathrm EMC}-{8m\0\g}=
\int_0^\L{dk\02\pi}\sqrt{k^2+m^2}\d'(k)+{4\0\g m}\d m^2\equiv 0
\ee
when using a strict energy cutoff.

On the other hand, invoking lattice regularization in a finite box
and comparing an equal number of modes in each sector yields
\bel{DMSMNC}
\D M_s^{\mathrm MNC}=
{1\02}\( \[ \sum_{-N}^{-1}+\sum_1^{N} \; \] \sqrt{ k^2_n+m^2}
-\sum_{-N}^N \sqrt{k'^2_n+m^2} \) +{4\0\g m}\d m^2
\ee
where we used that
in the presence of the soliton there is no solution 
of eqs.~(\ref{kdn}) and (\ref{dkSG})
with $n=0$. Its place is taken by the translational zero mode which
does not contribute to (\ref{DMSMNC}).
In the continuum limit we then obtain
\bel{MSGMNC}
\D M_s^{\mathrm MNC}=
-{m\02}-\int_0^\L{dk\02\pi}{d\sqrt{k^2+m^2}\0dk}
\d(k)+{4\0\g m}\d m^2=-{\L\d(\L)\02\pi}\to-{m\0\pi}
\ee
in agreement with Refs.\ \cite{DHN2,DV,Verw}.

The difference between (\ref{MSGMNC}) and (\ref{MSGEMC}) is seen to
have the same explanation as in the case of the kink, cf.\ (\ref{MKdiff}).

\subsection{Quantisation of the breather solution}

Besides the above soliton solutions, which are time-independent in
their rest-frame, one also knows exact time-dependent solutions
to the sine-Gordon field equations. We shall consider 
the famous ``breather'' or doublet solution,
\bea
\ph_\t(x,t)&=&{4\0\sqrt\g}\arctan\({\r\sin(2\pi t/\t)\0\cosh(2\pi\r x/\t)}\),
\quad {2\pi\0m}<\t<\infty\,,\nn
&&
\r\equiv \sqrt{\({m\t/2\pi}\)^2-1}\;,
\eea
which corresponds to a (topologically trivial) bound state of a
soliton and an anti-soliton.

Classically, its mass is given by $M_\t=2M_s\sqrt{1-(2\pi/m\t)^2}$.
In their monumental work, Dashen et al. \cite{DHNSG} have succeeded
to quantise this solution by an adaption of the semiclassical WKB method,
which leads to a refined Bohr-Sommerfeld-type quantization condition---the
DHN formula for the quantum bound states.

The DHN formula for the bound state energies reads \cite{DHNSG,R}
\beal{DHNf}
&&S_{cl}[\ph_\t]+S_{ct}[\ph_\t]+E_n \t(E_n)-\2\sum_i\n_i=2\pi n\\
&&E=-{d\0d\t}\(S_{cl}[\ph_\t]+S_{ct}[\ph_\t]-\2\sum_i\n_i\)
\eea
where $S_{cl}[\ph_\t]$ is the classical action of the breather solution
per period $\tau$, $S_{ct}$ the corresponding one from the counterterms,
and $\n_i$ are the so-called stability angles obtained from the solutions
of the linear stability equation
\bel{steq}
\(-{\6^2\0\6t^2}+{\6^2\0\6x^2}-V''[\ph_\t]\)\h_i(x,t)=0
\ee
with
\be
\h_i(x,t+\t)=e^{i\n_i}\h_i(x,t) \;.
\ee

As shown in Ref.\ \cite{DHNSG}, there are two solutions of (\ref{steq})
with vanishing stability angle and a continuum with
\be
\n_j=\t\sqrt{k_j^2+m^2}
\ee
where
\be
L k_j+\d(k_j)=2\pi j,\quad \d(k_j)=4 \arctan{\r(\t)\0k\sqrt{1+\r(\t)^2}}
\ee
when put in a finite box with periodic boundary conditions.

In Ref.\ \cite{DHNSG} (App.\ C), the evaluation of the sum over stability angles
was performed by means of lattice regularization in this finite box.
This leads to
\bea
\D S[\ph_\t]&\equiv&S_{ct}[\ph_\t]-\2\sum_i\n_i\nn&=&
-\t\[ \sum_{i=1}^N \(\sqrt{k_i^2+m^2}-\sqrt{k'^2_i+m^2}\)-m\]
+{16\pi\0\g m^2}\r(\t)\d m^2\;,\nn
\eea
where in counting the modes it is essential to note that $k_1=k_{-1}=0$
are degenerate modes.
In the limit $L\to\infty$ this becomes
\bea
\D S[\ph_\t]^{\mathrm MNC}&=&
\t \[ \int_0^\L {dk\02\pi}{d\sqrt{k^2+m^2}\0dk}
\d(k)+m \]+{16\pi\r\d m^2\0\g m^2}\nn&=&4(\r-\arctan(\r))=
-{\g\08\pi}S_{cl}[\ph_\t]\;.
\eea

Thus all the quantum corrections combine into a finite change of the
coupling constant
\be
\g\to\g'={\g\01-{\g\08\pi}}
\ee
and the bound state spectrum following from the DHN formula (\ref{DHNf})
is finally given by
\bel{SGbs}
E_n={16m\0\g'}\sin(n\g'/16),\quad n=1,2,\ldots < 8\pi/\g' \;.
\ee
The lowest-lying (bound) state is found to coincide with the ordinary
boson of the theory, $E_1=m$, and there is a series of bound-states
terminating at $E=16m/\g'$, which coincides with $2M_s^{\mathrm MNC}$
obtained by quantization in the soliton sector.

Repeating this calculation with a strict energy/momentum cutoff on
$\n/\t$ in place of
a lattice cutoff gives instead
\beal{DSEMC}
\D S[\ph_\t]^{\mathrm EMC}&=&-\t\int_0^\L {dk\02\pi}\sqrt{k^2+m^2}\d'(k)
+{16\pi\r\d m^2\0\g m^2}\nn&=&-{\g\08\pi}S_{cl}[\ph_\t]-4\r \;.
\eea

If taken literally, the beautiful result (\ref{SGbs}) is completely lost.
However, the extra term can be regarded as a finite mass shift that
would occur in addition to the replacement of $\g$ by $\g'$.
It can be absorbed by replacing $m$ by
\be m'={m\01-{\g\08\pi}}
\ee
in $S_{cl}[\ph_\t]$, which changes (\ref{SGbs}) to
\bel{SGbsEMC}
E_n^{\mathrm EMC}={16m\0\g}\sin(n\g'/16),\quad n=1,2,\ldots < 8\pi/\g' \;.
\ee
This indeed agrees with the quantum soliton mass as calculated before
with a strict energy cutoff, but now the lowest-lying state $E_1=
m'\not=m$. As we have remarked already, our renormalized value $m$
is the physical one as given by the pole of the boson propagator up
to terms of order $\g^2$. So we have lost the possibility to make
the identification of the lowest bound state with the
elementary boson. Moreover, the DHN result (\ref{SGbs})
has been checked thoroughly
against perturbative calculations of boson bound states \cite{DHNSG},
to which (\ref{SGbsEMC}) no longer fits.

Actually, the DHN formula has been derived by consistently neglecting
any mass renormalization except in $S_{ct}[\ph_\t]$ (cf.\
the discussion in Sect.\ III.C of Ref.\ \cite{DHNSG}). We could therefore
feel entitled to simply drop the extra term in (\ref{DSEMC}). This indeed
restores the result (\ref{SGbs}).

Now, however, the EMC result (\ref{MSGEMC}) for the mass of the quantum
soliton no longer fits nicely to the bound state spectrum. The highest
states would no longer disappear precisely when they can decay into a
soliton-antisoliton pair, but already before that.

From this we conclude that the regularization scheme
given by a strict energy/momentum cutoff is inadequate for calculating
quantum corrections in topologically nontrivial sectors,
whereas lattice regularization in a finite box leads to
consistent results.

\section{Quantum corrections to soliton masses in supersym\-metric theories}

Both of the two-dimensional models that we have considered above
have a supersymmetric extension given by a Lagrangian of the form\cite{DVFH}
\bel{Lss}
\CL=-\2\[ (\6_\m\ph)^2+U(\ph)^2+\5\psi\g^\m\6_\m\psi+U'(\ph)\5\psi\psi \]
\ee
where $\psi$ is a Majorana spinor, $\5\psi=\psi^{\mathrm T} C$. We shall use
a Majorana representation of the Dirac matrices with $\g^0=-i\t^2$,
$\g^1=\t^3$, and $C=\t^2$ so that $\psi={\psi^+\choose\psi^-}$ with
real $\psi^+(x,t)$ and $\psi^-(x,t)$.

The kink and the sine-Gordon systems are given by
\be
U_{\mathrm kink}(\ph)=\sqrt{\l\02}\(
\ph^2-\m_0^2/\l\), \quad
U_{\mathrm s\mbox{-}G}(\ph)={2m_0\0\sqrt\g}\sin{\sqrt\g\ph\02},
\ee
respectively.

Again we choose a minimal renormalization scheme as in (\ref{RSk}) and
(\ref{RSSG}), augmented now by
$Z_\psi=1$. Because of the additional fermionic contributions, we now
have
$$
\d m^2=m^2 {\g\08\pi}\int_0^\L {dk\0(k^2+m^2)^{1/2}} \eqno(\ref{RSSG}')
$$
in the sine-Gordon case.
One finds the same value for $\d m^2$ from the requirement that the bosonic
seagull graph for the fermionic two-point function vanishes. 
In fact, the complete counter-term contains now the following fermionic
terms
\be
\d \CL_{\mathrm ferm.}=-\2 m \5\psi\psi\(e^{\d m^2/2m^2}-1\)\cos{\sqrt\g\ph\02}
\ee
Just as in the purely bosonic case, both for the boson and the fermion
the physical mass is given by $m_P=m$ at one-loop order.

For the supersymmetric kink system
$$
\d\m^2=
{\l\02\pi}\int_0^\L {dk\0(k^2+m^2)^{1/2}}\;, \eqno(\ref{dm2k}')
$$
and the physical mass of the boson is given by
$$
m_P^2=m^2-{1\0\sqrt3}\l \;. \eqno(\ref{mP}')
$$
Also the physical mass of the fermion is given by (\ref{mP}$'$), in agreement
with supersymmetry.

In both models, the counterterm contribution to the soliton mass turns out
to be the same when expressed in terms of the minimally renormalized
boson mass, to wit,
\bel{dMss}
\d M={m\02\pi}\int_0^\L {dk\0(k^2+m^2)^{1/2}} \;.
\ee

In the formula for the leading quantum corrections to the soliton masses,
there is now also a fermionic contribution to the sum over eigenfrequencies
of fluctuations coming with a negative sign,
\bel{Msumbf}
M=M_{cl.}+\2\(\sum\o_B-\sum\o_B'\)-\2\(\sum\o_F-\sum\o_F'\)+\d M \;.
\ee
In the supersymmetric case, the two vacuum contributions cancel,
$\sum\o_B'-\sum\o_F'=0$, while the ones from the topologically nontrivial
sector clearly must not do so, for they have to combine with $\d M$ to
yield a finite result. This has been pointed out first by Schonfeld
\cite{Sch}, whereas it was overlooked in the early literature
on this subject \cite{dAdV,WO}.
As it turns out,
the crux is in the slightly different boundary conditions one has to
impose on bosonic and fermionic fluctuations.

The eigenvalue equations for the bosonic and fermionic normal modes read
\bea
&&\[-{d\0dx^2}+U'(\ph)^2+U(\ph)U''(\ph)\]\h_n=\o_{Bn}^2\h_n\;,\\
&&\[\g^1{d\0dx}+U'(\ph)\]u_n=i\o_{Fn}\g^0 u_n\;.
\eea
where we have written $\psi(x,t)=u(x)\exp(-i\o_F t)$.

With our choice of the Dirac matrices, we find the following two coupled
equations for $u^\pm$,
\bea
&&\[{d\0dx}+U'(\ph)\]u^+ + i\o u^- = 0 \label{upm1} \\
&&\[{d\0dx}-U'(\ph)\]u^- + i\o u^+ = 0 \label{upm2}
\eea

Acting with $(d/dx-U')$ on (\ref{upm1}), eliminating $u^-$ using
(\ref{upm2}), and substituting $\ph'=-U$ leads to
the same second-order equation for $u^+$ as for the bosonic fluctuations
$\eta$. Hence, the bosonic and fermionic eigenvalues are the same and
$u^+\sim \exp[i(kx\pm\2\d(k)]$ for $x\to\pm\infty$. From (\ref{upm2})
one then finds that $u^-(x) \sim \exp[i(kx\pm\2\d^-(k)]$
where 
\bel{thk}
\d^-(k)=\d(k)+\th(k)\equiv \d(k)-2\arctan{m\0k}
\ee
using that 
\bel{Upr}
U'(\ph(+\infty))=-U'(\ph(-\infty))=m
\ee
in both the kink and the sine-Gordon soliton background.

We now impose the boundary conditions \cite{KR}
\bel{fBC}
u^\pm(L/2) = u^\pm(-L/2) \;.
\ee
If one chooses $u^+ \sim \cos(kx\pm\2\d(k))$, this function satisfies
these boundary conditions, 
since it is even under $x \leftrightarrow -x$,
but the corresponding $u^-\sim\sin(kx\pm\2\d^-(k))$ is odd and thus must vanish
at $x=\pm L/2$ to satisfy its boundary condition. Conversely, if 
$u^-\sim\cos(kx\pm\2\d^-(k))$ and
$u^+ \sim \sin(kx\pm\2\d(k))$, the latter has to vanish at $x=\pm L/2$.
This yields the quantization conditions
\be
k^{(+)}_n L+\d(k^{(+)}_n)=2\pi n \quad\mbox{and}\quad
k^{(-)}_n L+\d(k^{(-)}_n)+\theta(k^{(-)}_n)=2\pi n
\ee
The solutions obtained by $k\to -k$ are not independent ones, since
$u^-$ follows from $u^+$ by (\ref{upm1}), and we may restrict $k$ to be
positive. 
Further, for $n=0$ there is no solution for $k^{(+)}$ 
satisfying (\ref{fBC}).
Hence, one half of the zero-point energy contributions from the
bosonic fluctuations
is cancelled by those half of the fermionic modes having
the same quantization condition as the bosonic ones, 
except for $n=0$, where $k_0^{(+)}$ is allowed by the boundary conditions
only for the bosonic modes. This leads to
\be
M=M_{cl}+\2\sum_{n\ge 0}\o(k_n^{(+)})-\2\sum_{n\ge 0}\o(k_n^{(-)})+\d M,
\ee
which has been obtained also in Ref.\ \cite{Sch} in a rather different
kind of analysis directly from the supersymmetry algebra.

However, Refs. \cite{Sch} and \cite{KR} disagree as concerns the
numerical result for the quantum correction to the soliton mass.
In Ref.\ \cite{KR}, 
the two sums were found
to exactly compensate with the counterterm contribution $\d M$ so
that there would be no correction to the soliton mass in a
minimal renormalization scheme. [The nonzero result given in Ref.\ \cite{KR}
is completely due to (\ref{mP}$'$).]
On the other hand, Schonfeld \cite{Sch} 
has obtained a finite difference.

We believe that both results are in error. The authors of Ref.\ \cite{KR}
(and also those of the later works of Ref.\ \cite{IM,CM,Y})
have implicitly used a strict energy/momentum cutoff for both sums, which in
analogy to the derivation of (\ref{MKEMC}) leads to
\be
M_{\mathrm EMC}=M_{cl}-{1\04\pi}\int_0^\L dk \sqrt{k^2+m^2}\th'(k)
+\d M=M_{cl}
\ee
upon inserting (\ref{thk}) and (\ref{dMss}).

On the other hand, 
starting from a lattice of finite extent, which 
in the previous sections we have found to be necessary
in order to obtain a consistent regularization, 
we obtain instead
\beal{Msusy}
M_{\mathrm MNC}&=&M_{cl}+\2\lim_{L\to\infty}\sum_{n=1}^N
\( \sqrt{k_n^{(+)2}+m^2}-\sqrt{k_n^{(-)2}+m^2}\)\nn
&=&M_{cl}+{1\04\pi}\int_0^\L dk {d\o\0dk}\th(k)+\d M=
M_{cl}+{\pi-2\04\pi}m
\eea
for both the supersymmetric kink and sine-Gordon soliton mass. Actually,
since in all these results $\d(k)$ drops out and $\th(k)$ is fixed by
(\ref{Upr}), the mass formula we have arrived at is universal to
soliton-bearing two-dimensional supersymmetric theories.

Comparing our result (\ref{Msusy})
with Schonfeld's \cite{Sch}, 
\be
M_{\mathrm Sch.}=M_{cl}-{m\02\pi},
\ee
who also evaluates
on a mode by mode basis, one finds that his different result comes
from his having included one mode more in the fermionic sum than in the bosonic
one (see eq.\ (2.45) of \cite{Sch}). 
Restoring equality in the number of modes would make his result
agree with ours, which by contrast features a positive mass correction. 
As we shall now discuss, a negative sign would be in conflict with
the renormalized Bogomolnyi inequality.

\section{The renormalized Bogomolnyi bound}

The supersymmetry algebra associated with models of the form (\ref{Lss})
reads
\bea
\{ Q^\pm,Q^\pm \} &=& 2 P_\mp,\\
\{ Q^+,Q^- \} &=& 2Z = 2 \int_{-\infty}^\infty U(\ph) \6_x \ph dx = 
2 \int_{\ph(-\infty)}^{\ph(+\infty)}U(\ph')d\ph',
\eea
where $Q=\int j^0 dx$, $j^\mu=-(\7\6+U)\g^\mu\psi$.

The central charge $Z$ depends, for a given model, only on $\ph(+\infty)$
and $\ph(-\infty)$, and is nonzero in topologically nontrivial sectors.
For the invariant mass squared $\CM^2=P_0^2-P_1^2=\2(P_+P_-+P_-P_-)$ 
one can derive
the Bogomolnyi bound \cite{WO}
\bel{Bb}
\CM^2=\2(\5QQ)^2+Z^2 \ge Z^2
\ee
which classically is saturated by the soliton solutions, $Z_{cl}=-M_{s,cl}$.

The renormalized operator
$\int U d\ph$ is given for the kink by $\sqrt{\l/2}({1\03}\ph^3-{\mu^2\0\l}\ph
-{\d\mu^2\0\l}\ph)$ with $\ph=\ph_K+\h$. We then obtain \cite{IM}
\bea
Z&=&Z_{cl}+\int_{-\infty}^\infty dx {d\0dx}\[\2U'(\ph_s) \< \h^2(x) \> \] 
+ \d Z \nn
&=&Z_{cl}+\2\[U'(+\infty)-U'(-\infty)\]{1\02\pi}\int_o^\L {dk\0\sqrt{k^2+m^2}}
+\d Z \equiv Z_{cl}
\eea
upon using 
\be
\d Z=\sqrt{\l\02}(-{\d\mu^2\0\l})\[\ph_K(+\infty)-\ph_K(-\infty)\]
\ee
Analogously
one finds $Z=Z_{cl}$ also for the supersymmetric sine-Gordon model.

Combined with the vanishing quantum correction to the soliton mass when
a simple energy/momentum cutoff is used, this result was taken \cite{IM}
to imply that the Bogomolnyi bound remains saturated by solitons even in the
quantum theory. For higher dimensions, a proof of this saturation at the
quantum level has been given by Witten and Olive \cite{WO}, but it ceases
to apply in two dimensions, because
``multiplet shortening'' does not occur. 
It has been conjectured by them to hold true
also in two dimensions, but this conjecture was motivated by the
results of Ref.\ \cite{dAdV} which have been corrected by Schonfeld \cite{Sch}
and Kaul and Rajaraman \cite{KR}. Only the latter result would indicate
a continued saturation of the Bogomolnyi bound, while the former is
inconsistent with the bound due to a negative soliton mass correction.
However, we believe to have shown that the regularization procedures used
respectively in Ref.\ \cite{Sch} and Refs.\ \cite{KR,IM,Y,CM} are inconsistent. 
Using a well-defined lattice regularization 
as proposed in Refs.\ \cite{DHN2,DHNSG}
we obtain a nonzero result for the quantum correction to the soliton mass
which is seen to be consistent with the inequality (\ref{Bb}) while
implying a loss of its saturation by solitons on the quantum level.

It would be very interesting to apply our methods 
also to the 3+1-dimensional $N=2$
supersymmetric systems with monopoles and to reinvestigate
the question of saturation of the Bogomolnyi bound there \cite{K2,IM2}.
Multiplet shortening only occurs if the supersymmetry algebra is not
modified at the quantum level, which can be decided by an extension
of our analysis.

\section{Epilogue: A failed rescue attempt}

After having come to the above conclusions, we learnt of a paper by
Uchiyama \cite{U1}, where our result (\ref{Msusy}) has been derived
in a somewhat different manner
directly from the supersymmetry algebra, but by using a mode number
cutoff (without discussing the
critical sensitivity on the particular UV regularization). However,
the initial conclusion \cite{U1} that the quantum soliton states
in two dimensions
do not saturate the Bogomolnyi bound has been withdrawn in a later
analysis \cite{U2}. In Ref.\ \cite{U2}, it was proposed that physical
observables such as the Hamilton operator
need to be restricted to a small fraction of the finite
box used in the quantization procedure in order to become less sensitive
to boundary conditions.

In particular, the Hamiltonian (but only when it is used to measure
the mass of the soliton state) is modified by
\be
H(f^2)=\int_{-L/2}^{L/2} dx f^2(x) {\cal H}
\ee
where $\cal H$ is the Hamiltonian density and $f(x)$ is defined by
\be
f(x)=\cases{0, &for $L/2 \ge |x| \ge (\ell+d)/2$\cr
            (\ell/2-|x|)/d+1/2, &for $ (\ell+d)/2 \ge |x| \ge (\ell-d)/2$\cr
            1, &for $ |x|\le (\ell-d)/2$\cr}
\ee
where $L\gg\ell\gg d\gg 1/m$.

This has the effect of changing all contributions proportional to $\hbar\o_n$
according to (see p.~130 of \cite{U2})
\be
\hbar\o_n \to \hbar\o_n {\int_{-L/2}^{L/2} dx f^2(x) \h_n^*(x)\h_n(x) \0
\int_{-L/2}^{L/2}\, dx\h_n^*(x)\h_n(x)},
\ee
where in the fermionic case one has to substitute the two-component
spinor $u_n$ for $\h_n$.
As a consequence,
the final result for the quantum soliton mass (\ref{Msusy}) gets
replaced by\cite{U2}
\bel{MsusyU}
M_{\mathrm U}=
M_{cl}+\lim_{d\to\infty}\lim_{\ell\to\infty}\lim_{L\to\infty}
{\ell-{1\03}d\0L}{\pi-2\04\pi}m=M_{cl}
\ee
(in the minimal renormalization scheme) and one thus gets rid of the
extra contribution that otherwise is responsible for the loss of the saturation
of the Bogomolnyi bound.

However, if this prescription is to be the correct one, one has to
redo also the calculations of the quantum corrections to the nonsupersymmetric
soliton masses. In the case of the ordinary kink, this has been done
in App. C of Ref. \cite{U2}. Here the result turns out to coincide
with the one we have obtained by using a strict energy/momentum
cut-off, eq.\ (\ref{MKEMC}).

It is a straightforward matter to perform the analogous calculation
for the sine-Gordon model.
This requires to calculate
\be
{\int_{-L/2}^{L/2} dx f^2(x) \h_n^*(x)\h_n(x) \0
\int_{-L/2}^{L/2} dx\,\h_n^*(x)\h_n(x)}=
{(\ell-{1\03}d)(k^2+m^2)/m-2\0L(k^2+m^2)/m-2}
\ee
for the fluctuations around the sine-Gordon soliton,
which modifies the result (\ref{MSGMNC}) according to
\bel{MSGU}
M_{\mathrm U}=
M_{cl}-\lim_{d\to\infty}\lim_{\ell\to\infty}\lim_{L\to\infty}
{\ell-{1\03}d\0L}{m\0\pi}=M_{cl} \;.
\ee
Here, again, we arrive at the
result we had obtained with an energy/momen\-tum cutoff,
eq.\ (\ref{MSGEMC}). But in this case we are getting into the very
same trouble that led us to discard the regularization by a strict
energy/momentum cutoff: the quantum
soliton mass no longer fits to the bound state spectrum obtained from
quantization of the breather solution. 

Notice however that we have been instructed to modify the Hamiltonian
in its role as observable only, but not when e.\ g.\ the fluctuation
spectrum is derived. We thus believe that the prescription of Ref.\ \cite{U2}
is inconsistent and that we have to come back to the previous conclusion of a
loss of saturation of the Bogomolnyi bound in two-dimensional
supersymmetric theories.

\section*{Appendix: Alternative boundary conditions}

In the purely bosonic case we have followed Refs.~\cite{R,DHN2,DHNSG} in
choosing periodic boundary conditions for the fluctuations about the
solitons. An alternative is to require that instead
\bel{vbc}
\eta(-L/2)=\eta(L/2)=0 \;.
\ee

This changes the quantisation relation (\ref{kdn}) to
$$
k_nL+ \d(k_n) = \pi n = k_n'L	\eqno(\ref{kdn}')$$
In the vacuum sector, (\ref{vbc}) excludes the constant mode, and in
the soliton sector, the solution with $n=0$.\footnote{Strictly speaking,
the $n=0$ is excluded also with periodic boundary conditions, because
its derivative is odd, but in the limit of large $L$ the latter vanishes
exponentially fast, and we have followed Refs.~\cite{R,DHN2,DHNSG} in
therefore not excluding it.} 
Eq.~(\ref{mnck}) has then to be replaced by
\beal{mnck2}
\D E_{\mathrm kink}&-&\D E_{\mathrm vac}=
{\hbar\02}{\sqrt3m\02}+{\hbar\02}\sum_{n=1}^{2N-2}\sqrt{k_n^2+m^2}
-{\hbar\02}\sum_{n=1}^{2N}\sqrt{k_n'^2+m^2}\nn
&=&{\hbar\02}{\sqrt3m\02}-\hbar m+\hbar\sum_{n=1}^{2N}
\(\sqrt{k^2_{n-2}+m^2}-\sqrt{k'^2_n+m^2}\)
\;.
\eea
Now 
$$
k_{n-2}L+[\d(k_{n-2})+2\pi]=\pi n=k'_nL \;,
$$
and the rest of the calculation turns out to be exactly the same as before.
Hence, both results for the mass of the soliton are unchanged.

In the supersymmetric case we have started from the assumption that
in the vacuum sector the bosonic and fermionic zero-point energies
cancel exactly. In order that this holds true mode by mode, we cannot
choose the boundary conditions for bosons and fermions independently.
With the choice of periodic boundary conditions for the bosonic fluctuations,
we have to adopt even boundary conditions for both $u^+$ {\em and} $u^-$.
(Antiperiodic boundary conditions for bosonic fluctuations would
require odd ones for the fermionic ones.)
Requiring instead that $\eta$ vanishes at the boundaries, we have to
require also that either $u^+$ {\em or} $u^-$
vanishes at the boundaries. In Ref.~\cite{Sch}
a further boundary condition has been considered: $\6_x\eta$ and one
of the spinor components vanishing at the boundaries, but this would
exclude the constant mode of the spinors without excluding the constant
mode of the bosonic sector, in violation of supersymmetry.

In Ref.~\cite{Sch}
\bel{Schbc}
u^-(L/2)=u^-(-L/2)=0 \quad \mbox{and} \quad
\eta(L/2)=\eta(-L/2)=0.
\ee
has been selected as the appropriate boundary conditions on grounds
that certain integrations by part are allowed, and in particular that
the Hamiltonian is time independent: $\dot H=\int\6_x [ \6_x\ph\6_0\ph-
\2\5\psi\g^1\6_x\psi]dx=0$. However, also our set (\ref{fBC}) allows
partial integration and leads to $\dot H=0$. It has the
virtue of being more symmetric in $u^+$ and $u^-$ and moreover guarantees
that all boundary terms drop from the supersymmetry algebra except
for the central charge \cite{U1}.

With the boundary conditions (\ref{Schbc}) we even find
that the standard semiclassical formula (\ref{Msumbf}) does not lead
to a finite result: the divergent contribution from the sum over
the difference of zero-point energies is twice the required value.
Alternatively,
with $u^+$ being set to zero at the boundaries, one instead finds
a vanishing contribution from this sum, which likewise does not
match the divergent contribution $\d M$. Indeed, the calculations
performed in Ref.~\cite{Sch} determine the quantum correction to the
soliton mass from the expectation value of $Q^2$ in the soliton sector.
From this a formula is obtained which differs from (\ref{Schbc}) by
having $1/4$ instead of $1/2$ in front of all the sums.

We therefore conclude that with the standard semiclassical formula 
(\ref{Msumbf}) the requirement of mode-by-mode matching in the vacuum
sector and ultraviolet finiteness in the soliton sector restricts
the choice to either periodic bosonic and even fermionic fluctuations
or antiperiodic bosonic and odd fermionic ones. In conformity with
most of the literature we have adopted the former in the present paper.

\subsection*{Acknowledgments}

AR wishes to thank 
Harald Skarke for useful discussions.

\end{document}